\documentclass[english,prd,superscriptaddress,nofootinbib,preprintnumbers,onecolumn,showpacs]{revtex4}
\usepackage[utf8]{inputenc}
\usepackage[english]{babel}
\usepackage{amsmath}
\usepackage{amsfonts}
\usepackage{amssymb}
\usepackage{epsfig}
\usepackage{graphics,psfrag,rotating}
\usepackage{graphicx}
\usepackage{dcolumn}
\usepackage{bm}
\usepackage{epstopdf}
\usepackage{color}
\usepackage[usenames,dvipsnames,svgnames]{xcolor}
\usepackage[T1]{fontenc}
\usepackage{multirow}
\usepackage{float}
\usepackage{comment}

\usepackage{subfigure}

\usepackage{enumitem}
\usepackage[colorlinks=true,
            linkcolor=red,
          urlcolor=gray,
            citecolor=blue]{hyperref}

\def\3nab{\tilde{\nabla}}

\def\be {\begin{equation}}
\def\ee {\end{equation}}
\def\ba {\begin{align}}
\def\ea {\end{align}}

\def\bc {\begin{center}}
\def\ec {\end{center}}
\def\case#1/#2{\frac{#1}{#2}}

\newcommand{\bver}{\begin{verbatim}}
\newcommand{\ever}{\end{verbatim}}

\newcommand{\bea}{\begin{eqnarray}}
\newcommand{\eea}{\end{eqnarray}}
\newcommand{\beaa}{\begin{eqnarray*}}
\newcommand{\eeaa}{\end{eqnarray*}}

\def\case#1/#2{\textstyle\frac{#1}{#2}}

\begin{document}

\title{Neutron Stars in frames of $R^{2}$-gravity  and Gravitational Waves}

\author{Artyom V. Astashenok\footnote{aastashenok@kantiana.ru}, Alexey S. Baigashov, Sergey A. Lapin}
\affiliation{I. Kant Baltic Federal University, Institute of
Physics, Mathematics and IT, Nevskogo st. 14, 236041 Kaliningrad,
Russia}



\begin{abstract}
The realistic models of neutron stars are considered for simple
$R+\alpha R^2$ gravity and equivalent Brance-Dicke theory with
dilaton field in Einsein frame. For negative values of $\alpha$ we
have no acceptable results from astrophysical viewpoint: the
resulting solution for spherical stars doesn't coincide with
Schwarzschild solution on spatial infinity. The mass of star from
viewpoint of distant observer tends to very large values. For
$\alpha>0$ it is possible to obtain solutions with required
asymptotics and well-defined star mass. The mass confined by
stellar surface decreases with increasing of $\alpha$ but we have
some contribution to mass from gravitational sphere appearing
outside the star. The resulting effect is increasing of
gravitational mass from viewpoint of distant observer. But another
interpretation take place in a case of equivalent Brance-Dicke
theory with massless dilaton field in Einstein frame. The mass of
star increases due to contribution of dilaton field inside the
star. We also considered the possible constraints on $R^{2}$ gravity from GW 170817 data.
According to results of Bauswein et al. the lower limit on threshold mass is $2.74^{+0.04}_{-0.01}$ $M_{\odot}$.
This allows to exclude some equations of state for dense matter. But in $R^2$ gravity the threshold mass increases for given equation of state with increasing of $\alpha$.
In principle it can helps in future discriminate between General Relativity and square gravity (of course one need to know equation of state with more accuracy rather than now).

\end{abstract}


\maketitle

\section{Introduction}
The general theory of relativity (GTR) is one of the most
carefully checked physical theories to date. Within the framework
of the GTR, it was possible to describe phenomena that can't be
explained using Newton's theory of gravity (perihelion precession
of planets, gravitational lensing), to predict a number of
effects, to explain the existence of an upper limit to the mass of
neutron stars. However, at the present moment the GTR is facing
the problem of ''dark energy'' in cosmology. Observational data of
type Ia supernovae and baryon acoustic oscillations in the cosmic
microwave background provide strong evidence that the universe is
expanding with acceleration. If one assumes that the universe is
filled with just matter and radiation, such cosmological dynamics
can't be explained. The accelerating expansion can be caused by a
substance with negative pressure (''dark energy'') but its
physical nature is unclear. It is also worthwhile to mention the
old problem of ''dark matter''. Possible candidates for
constituting ''dark matter'' (so called WIMPs --- weakly
interacting massive particles) are still not found using the Large
Hadron Collider.

Accepted by the majority in the scientific community, the
$\Lambda$CDM-model assumes that dark energy is nothing else but
the non-zero vacuum energy. The cosmological constant idea was
introduced already by A. Einstein with the purpose of constructing
a stationary solution for the cosmological equations. It turns out
the cosmological constant can also lead to the accelerating
expansion of the universe (exponentially with time). From a
phenomenological point of view, this model adequately describes
observational data but has some fundamental disadvantages. The
quantum field theory approach of describing gravity leads to the
theoretical vacuum energy value being greater than the
observational one by 120 orders of magnitude.

The discovery of the accelerating expansion of the universe
stimulated the search for models of gravity that can be used to
describe the cosmological evolution consistent with observations
without ``dark'' components (ref. \cite{Capozziello2002,
Capozziello2003, Odintsov2003, Carroll2004}). Models of modified
gravity are interesting in that it is possible to offer a unified
description of the cosmological acceleration and the early
inflation within their framework \cite{Odintsov2011,
Capozziello2010, Capozziello2011, Cruz2012}. Observational data
of type Ia supernovae or the cosmic microwave background
anisotropy can also be explained using these models
\cite{Capozziello2011} - \cite{Hwang2001}.

It is not possible to confirm or to disprove such theories using
only cosmological observational data. However, if the GTR just
approximately describes the real gravity, one can hope that
deviations from the GTR can be found in strong gravitational
fields \cite{Psaltis2008}. Nature has that kind of unique
``laboratories'' where our ideas about gravity can be tested.
These objects are neutron stars. The matter inside them is
compressed to the densities of $\sim 10^{15}\, \mbox{g/cm}^3$, and
the ``escape velocity'' near the surface is close to the speed of
light. Alternative theories of gravity must at least not
contradict the very fact of the existence of relativistic stars.
The more detailed analysis includes examinations of the
mass-radius relations, the inertial characteristics and the
rotation in these theories for neutron stars. Comparison of the
obtained results with the GTR results allows to determine the
possible ways of testing theories alternative to the GTR.

Parameters of neutron stars (mass, radius, moment of inertia etc)
depend decisively from equation of state (EoS). The (EoS) for
extreme dense matter in neutron stars is one of the puzzle of
modern astrophysics. The measurement of neutron stars masses can
be done with high precision by using post-Keplerian parameters.
From recent observations \cite{Demorest, Antoniadis} it follows
only that maximal mass of neutron stars is around $2M_{\odot}$.
This constraint on maximal mass ruled out many soft EoS of nuclear
matter mainly with hyperons. Unfortunately the determination of
neutron stars radii is more complicated task. Its values can be
obtained from X-ray spectra observation emitted by the atmosphere
of star. A large number of unknown parameters determine process of
emission and therefore estimations of neutron star radius from
such observations are different from each to other (see \cite{Ozel}
- \cite{Lattimer-16}). Finally one need to point that there are no
well-defined simultaneous measurements of mass and radius for
neutron stars. Therefore current observations give only weak
constraints on properties of neutron stars. Although many EoS with
maximal mass limit $M_{max}<2M_{\odot}$ are considered now as
unrealistic inaccuracy in the knowledge of exact dependence
between mass and radius ($M-R$ diagram) is very large for
discrimination between dozens of another realistic EoS.

The simple $R$-squared theory of gravity was considered as viable
alternative to GTR for description of neutron stars in many paper.
Initially the perturbative approach was used.  The scalar
curvature $R$ is defined by Einstein equations at zeroth order on
the small parameter, i.e. $R \sim  T$, where $T$ is the trace of
energy-momentum tensor. This approach is  applied to constructing
of neutron star models in $f(R)=R+\alpha R^2+\beta R^3$ and
$f(R)=R+\alpha R^2(1+\gamma \ln R)$ gravity also in
\cite{Cooney2010} - \cite{Astashenok2013}.

In modified $f(R)$ gravity model with cubic and quadratic terms,
it is possible to obtain neutron stars with $M\sim 2M_{\odot}$ for
simple hyperon equations of state (EoS) although the soft hyperon
equation of state is usually treated as non-realistic in the
standard General Relativity \cite{Astashenok2014}. The possible
signatures of modified gravity in neutron star astrophysics also
can include existence of neutron stars with extremely magnetic
fields \cite{Cheoun2013, Astashenok2015}.

The paper is organized as follows. We considered realistic models
for simple $R+\alpha R^2$ gravity and equivalent Brance-Dicke
theory with dilaton field in Einsein frame. Basic equations and
numerical scheme are presented in Section 2. The results of
calculation including mass profile, mass-radius diagram are given
in the next section. One can found that negative values of
$\alpha$ there is no acceptable result. The gravitational mass
infinitely grows with distance. For positive $\alpha$ it is
possible to obtain models with well-defined star mass. The mass
confined by stellar surface decreases with increasing of $\alpha$
but we have some contribution to mass from gravitational sphere
appearing outside the star. The resulting effect is increasing of
gravitational mass from viewpoint of distant observer. Section 4
is devoted to possible discrimination between $R^2$ gravity and
General Relativity in light of recent detection of gravitational
waves from merging neutron stars (object GW170817). We use
results from Section 4 but in another interpretation namely in
frames of equivalent Brance-Dicke theory with massless dilaton
field in Einstein frame.  In such theory the mass of star increases
due to contribution of dilaton field inside the star. Such
interpretation is more clear for merging of neutron stars. We can assume that
merging of neutron stars can be considered in frames of General Relativity.
and therefore apply results for threshold masses obtained in GTR.
For given EoS the parameters of neutron stars in
modified gravity change in comparison with GTR and as consequence the
threshold mass increases with increasing of $\alpha$.

\section{Modified Tolman-Oppenheimer-Volkoff equations for $f(R)=R+\alpha R^2$ gravity}

The action for a simple $f(R)$-gravity model can be written as
(from now on, we use system of units $G=c=1$):
\begin{equation}\label{1}
{S}=\frac{1}{16\pi}\int {\rm d}^4 x \sqrt{-g}\,[R+f(R)],
\end{equation}
where the Einstein-Hilbert action, which is proportional to the
scalar curvature $R$, was explicitly expressed. Here $f(R)$ is a
real differentiable function of the scalar curvature.

Therefore the gravitational equations can be written as:
\begin{eqnarray}\label{2}
R_{\mu\nu}-\frac{1}{2} \, R \, g_{\mu \nu} = \frac{1}{1+f_{R}} \, [- 8 \pi \,T_{\mu\nu}-\nabla_{\mu}\nabla_{\nu}f_{R}\nonumber\\
+\,g_{\mu\nu}\,\nabla^{\alpha}\nabla_{\alpha}f_{R}+\frac{1}{2}(f(R)-R\,f_{R})\,g_{\mu\nu}].
\end{eqnarray}
Here $f_{R}\equiv{\rm d}f(R)/{\rm d}R$, and $T_{\mu\nu}$ are the
components of the energy-momentum tensor.

As applied to compact non-rotating stars, the solution of the
obtained equations should be sought in terms of the spherically
symmetric metric:
\begin{equation} \label{Metric}
{\rm d}s^{2}=B(r)\,{\rm d}t^{2}-A(r)\,{\rm d}r^{2}-r^{2}({\rm
    d}\theta^{2}+\sin^{2}{\theta}\,{\rm d}\phi^{2}).
\end{equation}
For unknown functions of the radial coordinate $A$ and $B$, and
for the scalar curvature $R$, we have the equations (\ref{eqAr} -
\ref{eqRr}) on the assumption that the energy-momentum tensor is
diagonal $T^{\mu}_{\nu}=\mbox{diag}(-\rho,p,p,p)$:
\begin{eqnarray} \label{eqAr}
A'&=&\frac{2rA}{3(1+f_{R})}\left[8 \pi A(\rho+3p)+\frac{A}{2}R-\frac{3B'}{2rB}+Af(R)\right.\nonumber\\
&&\left.-\,f_{R}\left(\frac{A}{2}R+\frac{3B'}{2rB}\right)-\left(\frac{3}{r}+\frac{3B'}{2B}\right)f_{2R}\,R'\right]\,,
\end{eqnarray}
\begin{eqnarray} \label{eqBr}
B''&=&\frac{B'}{2}\left(\frac{A'}{A}+\frac{B'}{B}\right)+\frac{2A'B}{rA}+\frac{2B}{(1+f_{R})}\left[- 8 \pi Ap\right.\nonumber\\
&&\left.-\frac{A}{2}R+\left(\frac{B'}{2B}+\frac{2}{r}\right)f_{2R}\,R'-\frac{A}{2}f(R)\right],
\end{eqnarray}
\begin{eqnarray} \label{eqRr}
R''&=&R'\left(\frac{A'}{2A}-\frac{B'}{2B}-\frac{2}{r}\right)-\frac{A}{3f_{2R}}\left[8 \pi(\rho-3p)\right.\nonumber\\
&&\left.-(1-f_{R})R-2f(R)\right]-\frac{f_{3R}}{f_{2R}}\,R'^{2}.
\end{eqnarray}
Here $\rho$ and $p$ are the density and the pressure of the matter
respectively, and the functions $f_{2R}$ and $f_{3R}$ are the
second and the third derivatives of the function $f(R)$ with
respect to the scalar curvature. The relationship between the
pressure and the density is given by the chosen equation of state.
The other equation can be derived from the Bianchi identity:
\begin{eqnarray}
p'&=&-\frac{\rho+p}{2}\,\frac{B'}{B}.
\end{eqnarray}
The resulting system of the differential equations can be solved
if the boundary conditions (at the centre of the star and at
infinity) are specified. Further, we will consider in detail the
simple case when $f(R)=\alpha R^2$, where $\alpha$ is a parameter.

We also give description of our task in terms of scalar-tensor
theory \cite{Kokkotas, Kokkotas-1}. One can consider the
equivalent Brans-Dicke theory with following coupling between
scalar field $\Phi=1+df(R)/dR$ and curvature $R$:

\be S_{g}=\frac{1}{16\pi}\int d^{4} x \sqrt{-g}\left(\Phi R -
U(\Phi)\right). \ee

\noindent Here $U(\Phi)=Rf'(R)-f(R)$ is potential of scalar field.
In Einstein frame under conformal transformation
$\tilde{g}_{\mu\nu}=\Phi g_{\mu\nu}$ we have
 \be S_{g}=\frac{1}{16\pi}\int d^{4} x
\sqrt{-\tilde{g}}\left(\tilde{R}
-2\tilde{g}^{\mu\nu}\partial_{\mu}\phi\partial_{\nu}\phi-4V(\phi)\right).
\ee
Here redefined scalar field $\phi=\sqrt{3}\Phi/2$ and
potential in $V(\phi)=\Phi^{-2}(\phi)U(\Phi(\phi))/4$ are
introduced.

For spacetime metric we use the same form as in the case of $f(R)$
gravity (with redefined metric functions $\tilde{A}$,
$\tilde{B}$): \be \label{metric2} d\tilde{s}^{2}=\Phi
ds^{2}=\tilde{B}^2(\tilde{r})
dt^{2}-\tilde{A}^{2}(\tilde{r}){d\tilde{r}}^{2}-\tilde{r}^2d\Omega^2,
\ee where $\tilde{r}^2=\Phi r^{2}$, $\tilde{B}^2=\Phi B^2$. From
relation
$$
\Phi A^2 dr^{2}=\tilde{A}^2 d\tilde{r}^{2}
$$
one can obtain the following link between $\tilde{A}$ and $A$ (for
$f(R)=R+\alpha R^2$ gravity):
$$
\tilde{A}(r)=A(r)\left(1+\frac{r}{2}\frac{d\ln\Phi}{dr}\right)^{-1}=A(r)\left(1+\frac{\alpha
r}{1+2\alpha R(r)}\frac{dR(r)}{dr}\right)^{-1}.
$$
Therefore the mass parameter $\tilde{m}(r)$ is
 \be
\tilde{m}(r)=\frac{r}{2}\left(1-A^{-1}\left(1+\frac{\alpha
r}{1+2\alpha R}\frac{dR}{dr}\right)^2\right) \ee For functions
$\tilde{A}({\tilde{r}})$ and $\tilde{B}(\tilde{r})$ we have the
following equations (the tildes are omitted for simplicity): \be
\label{TOV1-1} \frac{1}{r^2}\frac{d}{dr}\left
[r\left(1-A^{-1}\right)\right]=8\pi
e^{-4\phi/\sqrt{3}}\rho+e^{-2\lambda}\left(\frac{d\phi}{dr}\right)^{2}+V(\phi),
\ee \be \label{TOV2-1} \frac{1}{r} \left[
A^{-1}\frac{dB}{Bdr}-\frac{1}{r}\left(1-A^{-1}\right)\right] =8\pi
e^{-4\phi/\sqrt{3}}p+e^{-2\lambda}\left(\frac{d\phi}{dr}\right)^{2}-V(\phi),\ee
These equations are nothing else than ordinary TOV equations with
redefined density and pressure and additional density and pressure
of scalar dilaton field.

The hydrostatic equilibrium condition is
\begin{equation}\label{hydro-1}
    \frac{dp}{dr}=-\frac{\rho
    +p}{2}\left(\frac{dB}{dr}-\frac{2}{\sqrt{3}}\frac{d\phi}{dr}\right).
\end{equation}
For scalar dilaton field we have equation equivalent to equation
for scalar curvature in in $f(R)$ theory:

\be \label{TOV3-1} \square
\phi+\frac{dV(\phi)}{d\phi}=-\frac{4\pi}{\sqrt{3}}
e^{-4\phi/\sqrt{3}}(\rho-3p). \ee

Here $\square$ is D'Alamber operator in metric (\ref{metric2}).
The potential of scalar field in considered case of $R^2$-gravity
is \be \label{V100}
V(\phi)=\frac{1}{4\alpha}\left(1-e^{-2\phi/\sqrt{3}}\right)^2. \ee

The unknown parameters (the mass and the radius) are dependent on
the density and the pressure in the centre of a neutron star $p(0)
= p_c, \ \rho(0) = \rho_c$. We considered a representative set of
equation of states. Firstly the well-known APR EoS \cite{APR} is
obtained from three-nucleon potential and Argonne 18 potential
with UIX potential. For completeness, we have also considered the
SLy EoS \cite{SLy}, \cite{SLy-4} obtained from many body calculations
with simple two-nucleon potential. As an example of EoS based on
relativistic mean-field (RMF) calculations we take the GM1 model
firstly considered by Glendenning and Moszkowski \cite{GM}. We
included into consideration the realistic EoS proposed recently by
\cite{MYN}. The maximal mass for GM1 EoS is below  than limit
established of observations (for General Relativity) but for
$\alpha\sim 10$ we found that maximal mass is around two solar.

For the metric function $A(r)$, the following condition must be
met at large distances:
$$
A(r)=\left(1-\frac{2M}{r}\right)^{-1},
$$
where $M$ is the gravitational mass of an object. According to
this, we choose $A(0)=1$ as the condition in the centre
\cite{Resco2016}. The function $B(r)$ is included in the equations
as $B'/B$ and $B''/B$, so the solution of the system does not
depend on the value of $B$ in the centre of a star. The condition
$B(r)\rightarrow 1$ must be met at an infinite distance from the
star ($ \rightarrow \infty$). One can solve the equations for an
arbitrary value of $B(0)$, find the solution of $B(r)$ for
sufficiently large distances, extrapolate the asymptotics of the
solution for $r \rightarrow \infty$, and hence obtain the value of
$B_\infty$. Thus, the solution has the required asymptotics
$B\rightarrow 1$, if the initial value is given by
$\tilde{B}(0)=B(0)/B_\infty$.

The regularity of the equations at $r=0$ demands that for the
derivative $B'(r)$ at $r=0$ the condition $B'(0)=0$ is met. The algorithm for finding condition for the scalar
curvature in the centre was developed using the bisection method.
When the surface of the star is reached, the function $R(r)$ must
have a positive value and then decrease rapidly while
asymptotically approaching zero. The values of $R(0)$ were being
found in the range $-100 R_{0} < R < 100 R_0$, where $R_0 = 8 \pi
(\rho(0) - 3 p(0))$ is the value of the curvature in the GTR.

Integration of the equation system was performed using the
Runge--Kutta--Merson fourth-order method.

It should be noted that when the surface of the star is reached
(that is, any of the conditions $\rho(r) < 0$ or $p(r) < 0$ is
met) the system is simplified to the three equations (\ref{eqAr}),
(\ref{eqBr}), (\ref{eqRr}) with zero density and pressure.

\section{Results of calculations for $R^2$-gravity}

Let us further consider negative and positive values of $\alpha$.

\begin{figure}[h!]
\centering
\includegraphics[width=0.65\textwidth]{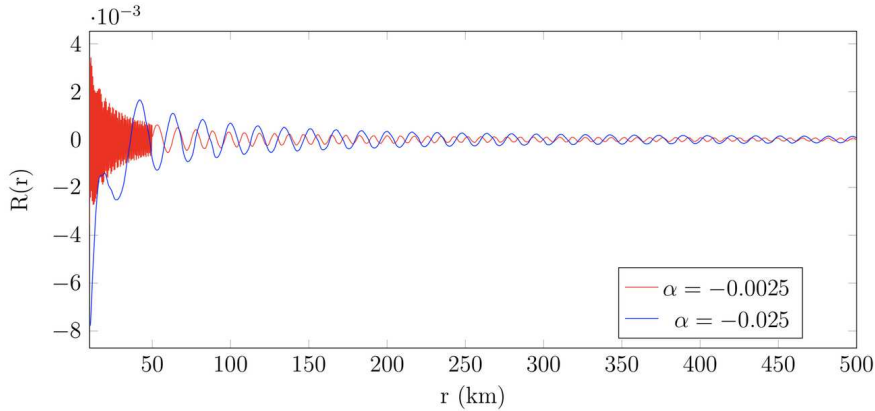}
\caption{{The scalar curvature $R$ as a function of the radial
coordinate for the star with the central density $\rho = 3 \cdot
10^{14} \, \mbox{g/cm}^3$, obtained for APR EoS. The scalar
curvature $R$ hereafter is given in units of $r_g^{-2}$ and parameter $\alpha$ in $r_g^{2}$, where
$r_g = \sqrt{G \, M_{\odot} / c^2}$.}}
\label{Fig:1}
\end{figure}

1) $\alpha<0$. In this case, the scalar curvature outside the star
oscillates (Fig.\ref{Fig:1}). The value of the scalar curvature in the
general theory of relativity $R(0) = 8 \pi (\rho(0) - 3 p(0))$ was
chosen as the boundary condition for the scalar curvature in the
centre of the star. Analysis shows that the qualitative form of
the solution for the scalar curvature outside the star depends
weakly on the value of $R(0)$.

\begin{figure}[h!]
\centering
\includegraphics[width=0.40\textwidth]{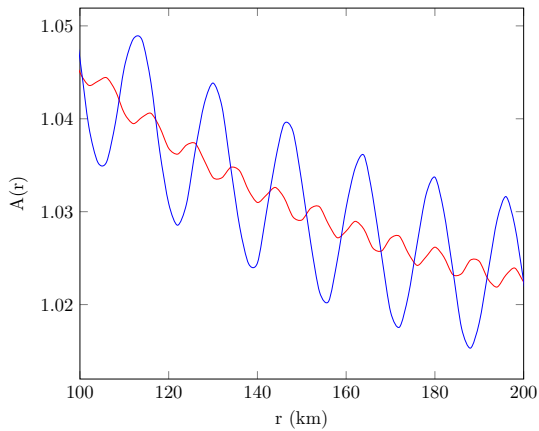}
\caption{{The function $A(r)$ as a function of the radial
coordinate in narrow interval for parameters as on previous
figure.}}
\label{Fig.3}
\end{figure}

\begin{figure}[h!]
\centering
\includegraphics[width=0.65\textwidth]{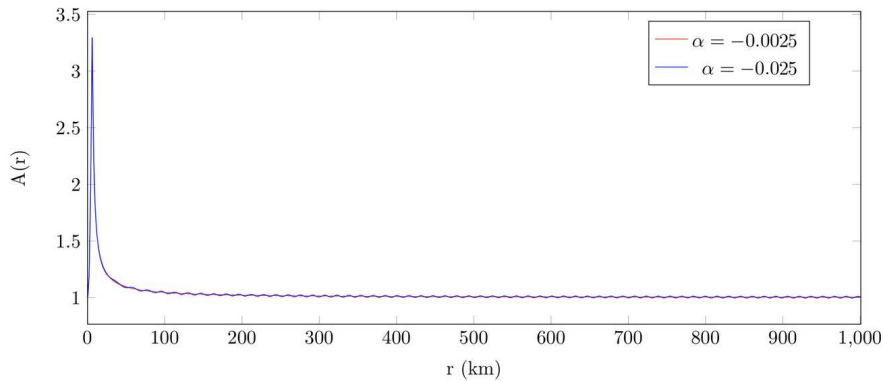}
\caption{{The function $A(r)$ as a function of the radial
coordinate for parameters as on Fig. 1.}}
\label{Fig.2}
\end{figure}

On the assumption that the solution for the functions $A(r)$ and
$B(r)$ at large distances tends to the Schwarzschild solution, one
can expect that
$$
A(r) \rightarrow \left(1-2M(r)/r\right)^{-1}, \quad r \rightarrow
\infty .
$$
However, the function $A(r)$ at large distances oscillates under
the law (see Fig.\ref{Fig.2} and Fig.\ref{Fig.3})
$$
A(r)\approx 1+A_{0}(r)\sin\left(\sqrt{6|\alpha|} \cdot r\right)
,\quad A_0(r)<<1,
$$
where $A_0(r)$ is a decreasing amplitude of the oscillation. This
makes it impossible to determine the gravitational mass using the
form of the solution for $A(r)$. One can use the solution for
$B(r)$. The function $B(r)$ approaches some constant value from
below as $r \rightarrow \infty$ so that $B(r)<B_{\infty}(r)$.

The main results of our calculations consist of the following:

1. For the large negative values $\alpha$, a significant increase
of gravitational mass is observed outside the star.

2. For the small negative values of $\alpha$, the mass function
undergoes a small smooth growth at large (in comparison with the
size of the star) distances. But in any case the gravitational mass grows infinitely.

\begin{figure}[h!]
\includegraphics[width=0.90\textwidth]{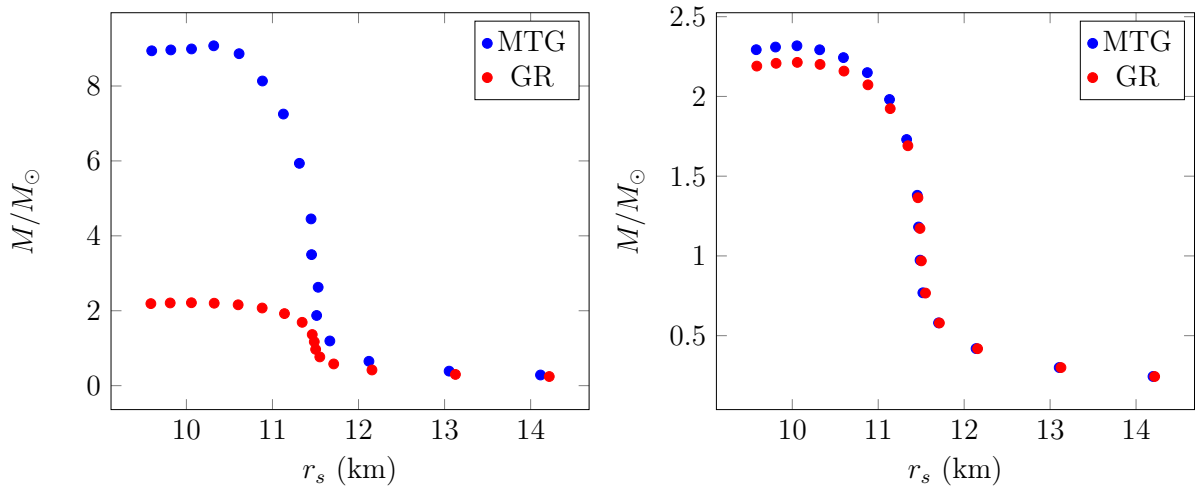}
\caption{{Mass-radius relation for APR EoS. On the left: $\alpha =
- \, 0.025$. On the right: $\alpha = - \, 0.0025$. The integration was performed to the distance of
$10^5 \ \mbox{km}$. Hereafter $r_s$ is a star radius.}}
\label{Fig.6}
\end{figure}

\begin{figure}[h!]
\includegraphics[width=0.90\textwidth]{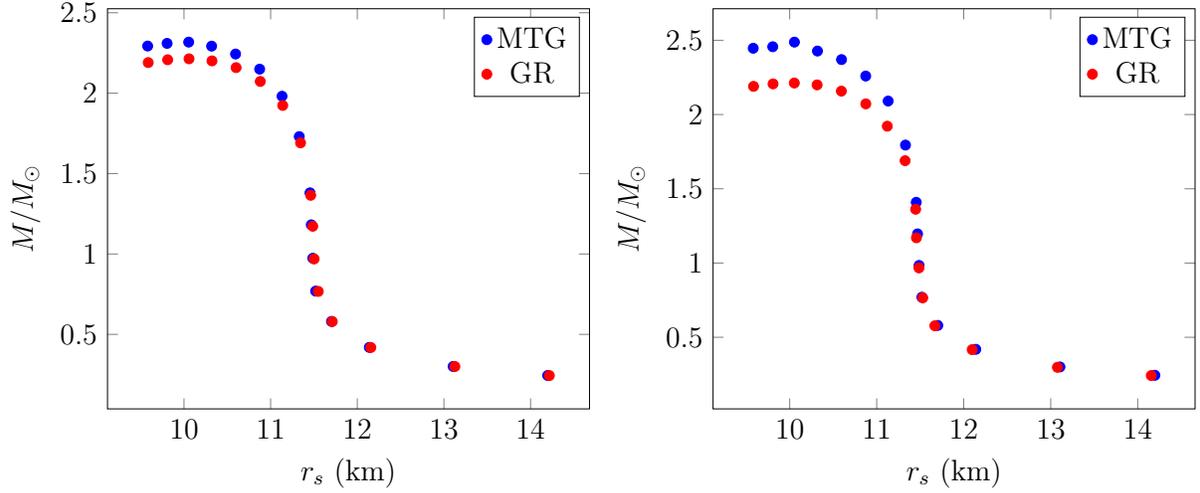}
\caption{{The same as on Fig. 4. On the left: $\alpha = -  \,
0.0025$, the integration was performed to the
distance of $10^5 \, \mbox{km}$. On the right: $\alpha = -  \,
0.0025$, the integration was performed to the
distance of $2.5\times 10^5 \, \mbox{km}$}.}
 \label{Fig.7}
\end{figure}

We depicted the mass-radius relation for negative values of
$\alpha$ for case where integration of equations was performed to
the distance $10^5$ km and $2.5\times 10^5$ km (see
Fig.\ref{Fig.6}, Fig.\ref{Fig.7}).

As a result of the interpolation of the gravitational mass
function to infinity $(r \rightarrow \infty)$, the value of the
function tends to infinity. Thus it can be argued that the
solution of the gravitational equations can not asymptotically
approach the Schwarzschild solution.

\begin{figure}[h!]
\includegraphics[width=0.90\textwidth]{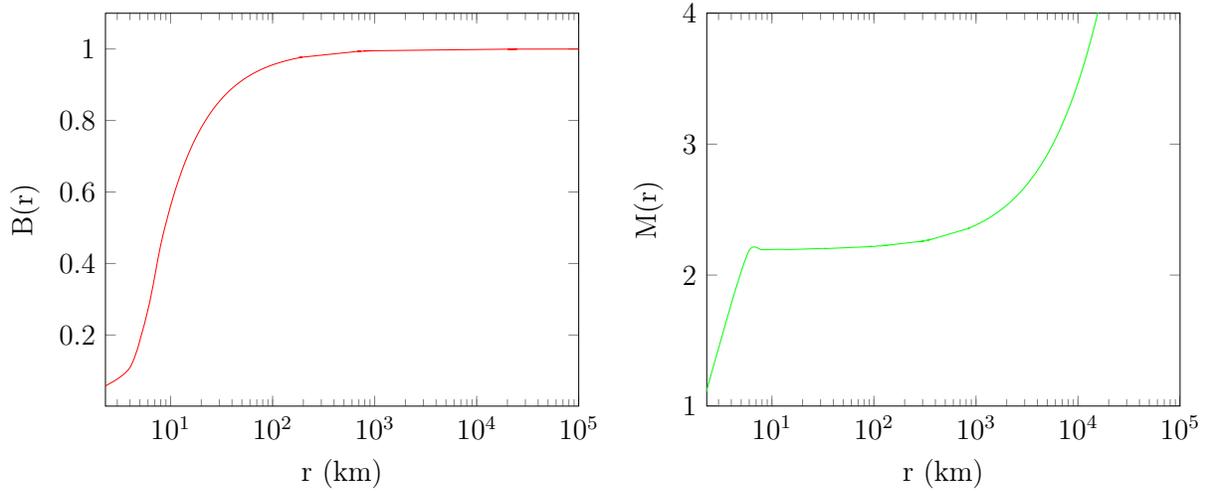}
\caption{{The profiles of functions $B(r)$ and $M(r)$ for $\alpha
= - 0.025$. The integration was performed to the
distance of $10^5 \ \mbox{km}$. The value of the density in the
centre of the star is $\rho = 3 \cdot 10^{14}  \, \mbox{g/cm}^3$}}
\label{Fig.8}
\end{figure}

\begin{figure}[h!]
\includegraphics[width=0.90\textwidth]{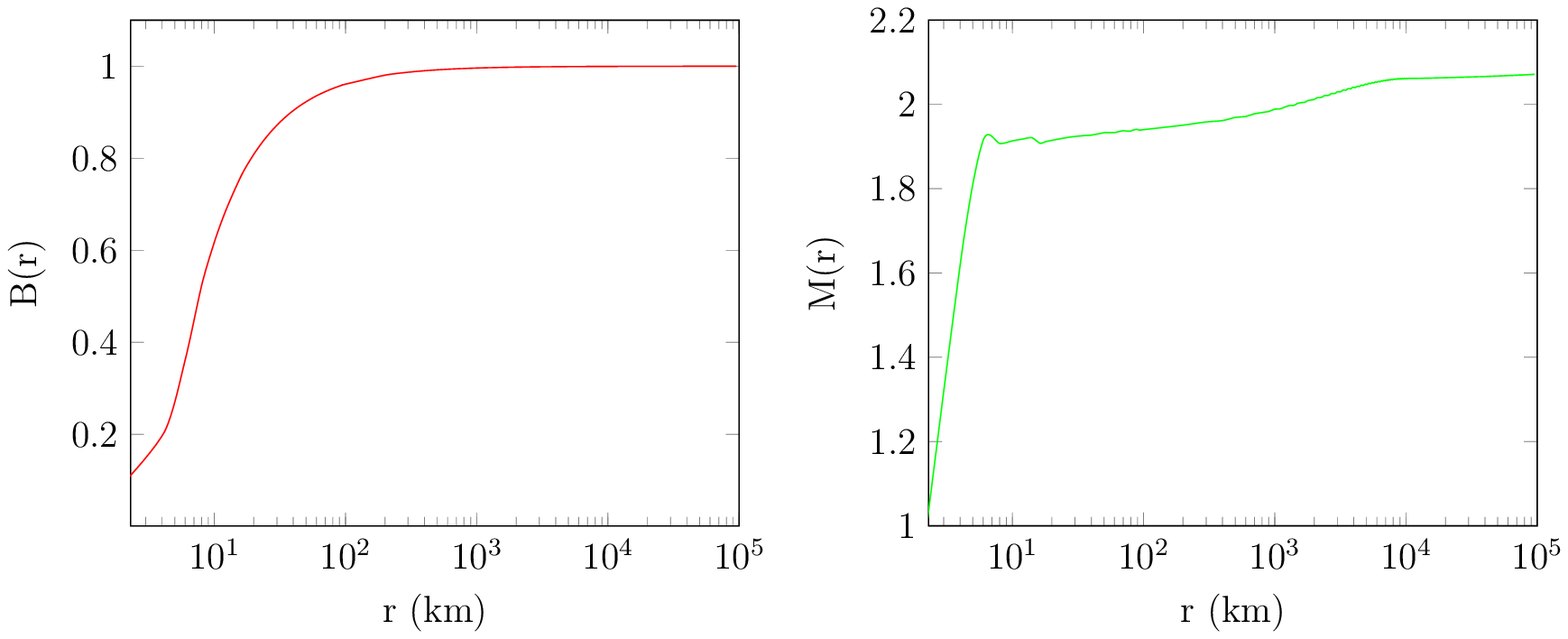}
\caption{The same as on Fig. 8 but for $\alpha = -  \, 0.0025$.}
\label{Fig.9}
\end{figure}

Although the mass function has an apparent kink and plateau for
the given equation of state with any initial values of the
pressure and the density (taken from the specified interval) in
the centre of the star, an increase in the mass with the distance
from the star is observed for the values of the parameter close to
$\alpha = - 0.025$. It is possible to reduce this
increase by decreasing the value of the parameter $\alpha$, yet it
is not possible to achieve a parallel alignment of the plateau for
any values of the parameter (see Fig.\ref{Fig.8}, Fig.\ref{Fig.9}).

Similar results have been obtained for other equations of state. This allows to conclude that the gravitational
mass of an object measured by a distant observer has an enormous
value. This obviously does not represent the observations.

As for the boundary condition for the scalar curvature $R$ in the
centre of the star, its changing does not lead to any qualitative
change in the behaviour of the system solution.

\begin{figure}[h!]
\centering
\includegraphics[width=0.65\textwidth]{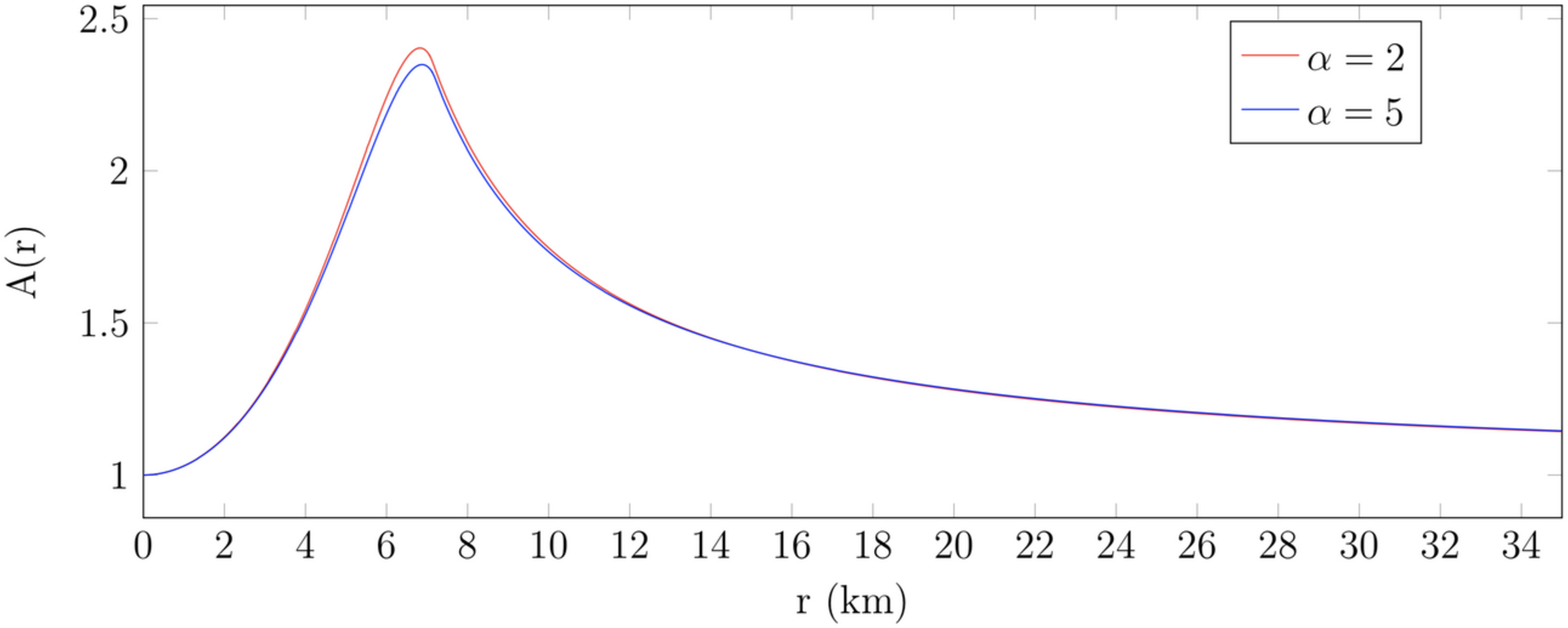}
\caption{{The same as on Fig. \ref{Fig.2} but for two positive
values of $\alpha$.}} \label{Fig.4}
\end{figure}

\begin{figure}[H]
\centering
\includegraphics[width=0.60\textwidth]{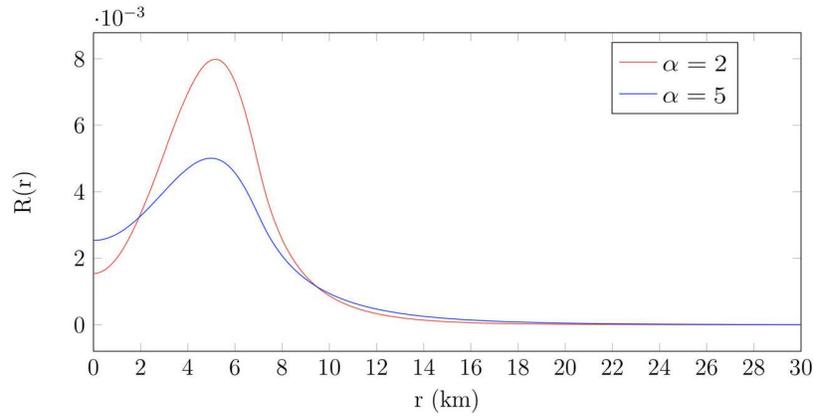}
\caption{{The same as on Fig. \ref{Fig:1} but for two positive
values of parameter $\alpha$.}}
 \label{Fig.5}
\end{figure}

2) $\alpha>0$. In the case of $\alpha>0$, the solution for the
scalar curvature and for the function $A(r)$ behaves differently
(Fig.\ref{Fig.4}, Fig.\ref{Fig.5}). The solution at large distances becomes asymptotic to
the Schwarzschild solution for only one value of the scalar
curvature in the centre. The mass of the star can be determined
using $A(r)$:
\begin{equation}
M(r) = \left( 1 - \frac{1}{A(r)} \right) \cdot \frac{r}{2}.
\end{equation}

\begin{figure}[h!]
\includegraphics[width=0.90\textwidth]{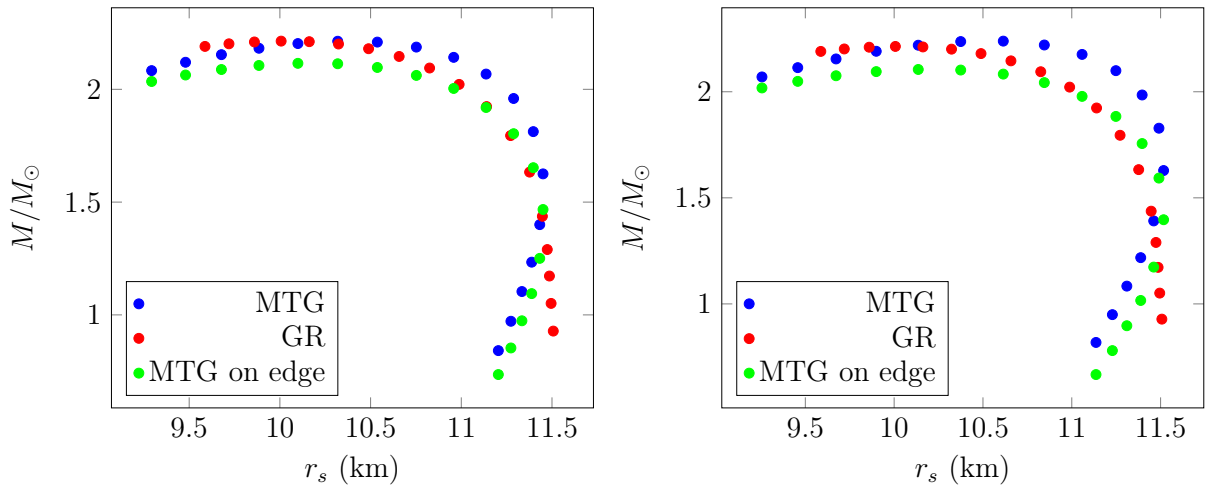}
\caption{{The mass-radius relation for $\alpha > 0$ using APR EoS.
On the left: $\alpha = 2$, the maximum mass is 2.20
$M_{\odot}$. On the right: $\alpha = 5$, the
maximum mass is 2.24 $M_{\odot}$. The integration was performed to
the distance of $50 \, \mbox{km}$.}}
\label{Fig.10}
\end{figure}

The mass-radius relation is given on Fig.\ref{Fig.10}.

\begin{figure}[h!]
\includegraphics[width=0.60\textwidth]{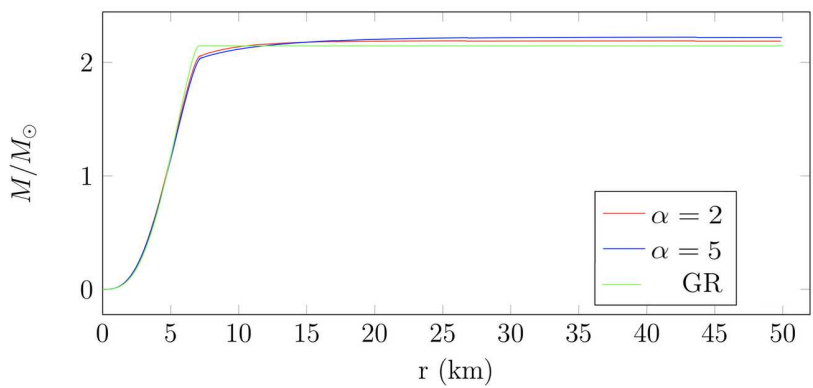}
\caption{{The profile of $m(r)$ for various values of $\alpha$.
The integration was performed to the distance of $50 \, \mbox{km}$.
The value of the density in the centre of the star is $\rho = 3
\cdot 10^{14}  \, \mbox{g/cm}^3$}. The profile of mass slowly
increases outside the star with distance.}
\label{Fig.11}
\end{figure}

Based on the obtained results, one can conclude that neutron stars
have larger radii at the given stellar mass in the considered
theory of gravity than in the GTR. However, the observed
gravitational mass is a sum of the mass of the star itself and the
mass of the ``gravitational sphere'' i.e. the region outside the
star where the scalar curvature decreases rapidly while
asymptotically approaching zero (see Fig.\ref{Fig.11}).

Considering the dependence of mass from radial coordinate in
Brance-Dicke theory in Einstein frame we have more clear picture.
The mass grows with $\alpha$ but the increasing of mass occurs
within star. Outside the star mass doesn't depend from radial
coordinate.

\section{Gravitational Waves and $R^2$ gravity}

Additional information about parameters of neutron stars can be
extracted from gravitational waves generated by possible merging
of neutron stars. First event of such type was detected recently
(GW170817, see \cite{LIGO} - \cite{Abbott-3}). The measured chirp
mass is $M_{c}=1.118^{+0.004}_{-0.002}$ $M_{\odot}$ and
corresponding total mass is $M_{tot}=2.74^{+0.04}_{-0.01}$ with
masses of components $M_{a}=1.36-1.60 M_{\odot}$ and
$M_{b}=1.17-1.36 M_{\odot}$. From analysis of spectrum of
gravitational waves performed by Abbott et al. \cite{Abbott-2}
follows that amplitude of tidal effects is relatively small and
therefore large radii of neutron stars are unacceptable.

From observations follows that electromagnetic radiation from
GW170817 included a weak gamma-ray burst kilonova emission from
the radioactive decay of the merger ejecta, and X-ray/radio
emission. If maximum mass is sufficiently large supramassive
neutron star appears after merging. This star should spin-down
before gravitational collapse into a black hole. As result
rotational energy is transferred into or Gamma Ray Burster (GRB) jet or kilonova (KN)
ejecta. Therefore it is possible estimate the upper limit of
neutron star with using gravitational wave signal and limits on
energy of GRB and KN from electromagnetic signal. This idea was
used in recent work \cite{Margalit}. According to this paper the
remnant appeared after merging  was not long lived, because of the
relatively low energy of the ejecta inferred from electromagnetic
wave data. This assumption gives upper limit on maximal mass of
neutron star as $2.17 M_{\odot}$ with 90\% confidence. Then
authors of \cite{Bauswein} suggested a powerful method to
constrain properties of neutron stars from total mass of GW170817.
Assumption that merger did not result in a prompt collapse
\cite{Kasen} allows to conclude that the radius $R_{1.6}$ and
$R_{max}$ of nonrotating NSs with a mass of $1.6$ $M_\odot$ and
maximal mass correspondingly should be larger than
$10.68^{+0.08}_{0.04}$ km and $9.60^{+0.04}_{-0.03}$ km. According
to hydrodynamical simulations \cite{Bauswein-2} the threshold mass
for given $R_{max}$ has a maximum for some mass. For larger radii
the threshold mass grows. Assuming some limit on threshold mass it
is possible to obtain upper limit on radius of star with maximal
mass or $1.6M_\odot$. Therefore combining various estimation one
can obtain the acceptable interval for radii.

According to estimations the mass transferred into electromagnetic
emission lies in interval $0.03M_{\odot}<M_{em}<0.05M_{\odot}$
\cite{Cowperthwaite, Chornock}. For hypothetical prompt collapse
one can put that threshold mass is
$$
M_{th}<M_{tot}=2.74^{+0.04}_{-0.01}M_{\odot}.
$$
Therefore total binary mass can be considered for delayed collapse
or no collapse scenario as lower limit for threshold mass. For
threshold mass authors of \cite{Bauswein} suggested following
approximation: \be \label{Mth}
M_{th}=\left(-3.606\frac{GM_{max}}{c^2 R_{1.6}}+2.38\right)M_{max}
\ee where $R_{1.6}$ is radius of nonrotating neutron star with
mass of $1.6M_{\odot}$ and $M_{max}$ means maximal value of NS
mass for given equation of state. Another approximation for
threshold mass can be defined via radius of stellar configuration
with maximal mass: \be \label{Mth2}
M_{th}=\left(-3.38\frac{GM_{max}}{c^2 R_{max}}+2.43\right)M_{max}
\ee For given equation of states (EoS) one can obtain the value of
$M_{th}$. These results can be applied for analysis of various
models of gravity. In any case we propose that deviations from GR
are sufficiently small and therefore the process of merging can be
described in frames of post-newtonian formalism and we can use the
approximation of threshold mass suggested in \cite{Bauswein}.
However the deviations from General Relativity affect on mass and
radius of star for given density in center and then threshold mass
for corresponding EoS should be vary.

As mentioned above for $f(R)=R+\alpha R^2$ gravity the mass confined
by stellar surface decreases with increasing of $\alpha$. The
total gravitational mass increases due to contribution of
gravitational sphere with nonzero scalar curvature appearing
outside the star. The radii of sphere is around around tens km for
realistic values of $\alpha$. Therefore
picture of merging is not clear. If we assume that part of
gravitational mass outside the star doesn't affect considerably on
process of merging one can conclude that threshold mass decreases.
In scalar-tensor theory in conformal gauge the situation is more
easy. Dilaton sphere outside the star appears but its contribution
to total gravitational mass is negligible. In principle one can
consider that deviation from General Relativity is equivalent to
some modification of equation of state for dense matter. Therefore
we can use approximation (Fig.\ref{Mth}), (Fig.\ref{Mth2}) for
threshold mass.

We investigated the varying of threshold mass for EoS mentioned above in a
case of scalar-tensor theory of gravity equivalent to
$f(R)=R+\alpha R^2$ gravity. The main result is increasing of
threshold mass with increasing of $\alpha$. One note that the difference between values calculated by using Eqs.
(\ref{Mth}) and (\ref{Mth2}) slightly increases with $\alpha$.
However even for General Relativity ($\alpha=0$) this difference
can exceed $0.05M_{\odot}$ (for Sly4 EoS). Increasing of threshold
mass for $\alpha=10^2$ is around of $0.12-0.14$ $M_{\odot}$
(according to (\ref{Mth})) in comparison with General Relativity.
For approximation (\ref{Mth2}) we have more interesting picture:
for SLy4 and APR EoS threshold mass increases by $\sim 0.25 $
$M_{\odot}$. In a case of MYN EoS increasing is $0.18$ $M_{\odot}$
and for GM1 we have again $0.14$ $M_{\odot}$.

Therefore from total mass of binary merger GW170817 we cannot in principle extract
rigid constraints on parameter $\alpha$. One can hope however that
in future the upper bound on threshold mass will be established.
For this one need to observe the event like GW170817 but with a
higher chirp mass and with evidence of prompt collapse. In this
case for any realistic EoS we can in principle estimate the upper
limit of $\alpha$.

\begin{table}[H]
\begin{centering}
\begin{tabular}{|c|c|c|c|c|c|c|c|c|c|c|c|c|c|c|c|}
  \hline
  $\alpha$  & \multicolumn{3}{c}{0} \vline & \multicolumn{3}{c}{10} \vline & \multicolumn{3}{c}{20} \vline & \multicolumn{3}{c}{50} \vline & \multicolumn{3}{c}{50}\vline\\
  \hline
  EoS      & $M_{max}$ & $R_{max}$ & $R_{1.6}$ & $M_{max}$ & $R_{max}$ &
        $R_{1.6}$ & $M_{max}$ & $R_{max}$ & $R_{1.6}$ & $M_{max}$ & $R_{max}$ & $R_{1.6}$ & $M_{max}$ & $R_{max}$ &
        $R_{1.6}$ \\
        \hline
  APR   & 2.23 & 10.01 & 11.29 & 2.23 & 10.46 & 11.47 & 2.25 & 10.60 & 11.58 & 2.29 & 10.86 & 11.73 & 2.31  & 10.96 & 11.87  \\
  GM1   & 1.93 & 12.12 & 13.66 & 1.96 & 12.17 & 13.62 & 1.97 & 12.27 & 13.68 & 2.00 & 12.44 & 13.81 & 2.03 & 12.56 & 13.93  \\
  SLY4  & 2.05 & 9.97 & 11.50 & 2.09 & 10.41 & 11.67 & 2.11 & 10.52 & 11.78  & 2.15 & 10.84 & 11.94 & 2.17 & 10.90 & 12.07   \\
  MYN   & 1.95 & 11.34 & 12.76 & 2.00 & 11.54 & 12.81 & 2.02 & 11.59 & 12.90 & 2.06 & 11.79 & 13.04 & 2.08  & 11.94 & 13.17  \\

\hline
\end{tabular}
\caption{The parameters of neutron stars (maximal mass, radius for
maximal mass configuration and radius $R_{1.6}$ for star with
$M=1.6M_{\odot}$) for various EoS in GTR and scalar-tensor gravity with potential (\ref{V100})
for some $\alpha$. The
radii are given in km.} \label{Table2}
\end{centering}
\end{table}

\begin{table}[H]
\begin{centering}
\begin{tabular}{|c|c|c|c|c|c|}
  \hline
  $\alpha$  & 0  & 10 & 20 & 50 & 100 \\
  \hline
  APR   & 2.97 (2.94) & 3.00 (3.05) & 3.03 (3.09)  & 3.07 (3.16) & 3.11 (3.19)    \\
  GM1   & 3.14 (3.16) & 3.17 (3.19) & 3.18 (3.21)  & 3.22 (3.26) & 3.26 (3.30)    \\
  SLy4  & 2.94 (2.88) & 2.98 (2.99) & 3.01 (3.02)  & 3.06 (3.10) & 3.09 (3.12)    \\
  MYN   & 3.06 (3.07) & 3.10 (3.13) & 3.13 (3.15)  & 3.17 (3.21) & 3.20 (3.25)    \\

\hline
\end{tabular}
\caption{Threshold masses in GTR and for scalar-tensor theory
calculated from Eqs. (\ref{Mth}) and (\ref{Mth2}) (in brackets)
correspondingly.} \label{Table2}
\end{centering}
\end{table}

\begin{figure}
\begin{center}
\includegraphics[scale=1]{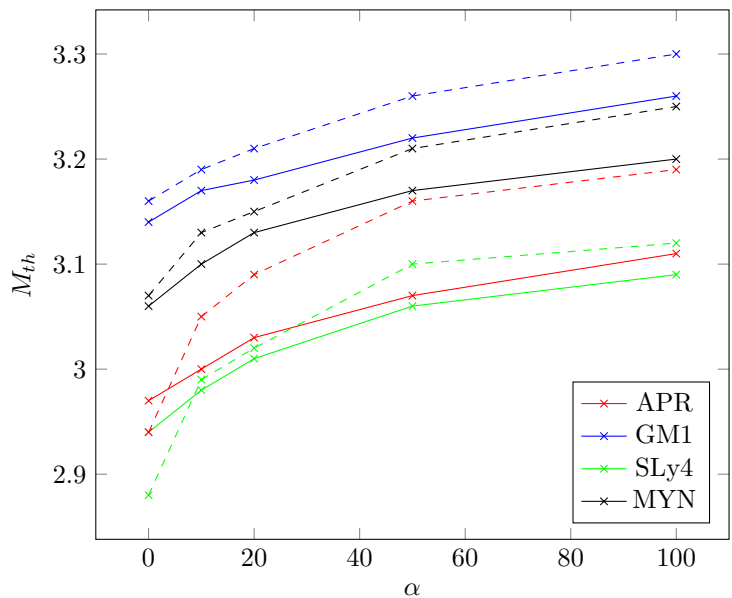}\\
\caption{Threshold masses for various EoS calculated according to
Eqs. (\ref{Mth}) (solid) and (\ref{Mth2}) (dashed) as function of
parameter $\alpha$.} \label{Fig12}
\end{center}
\end{figure}

\section{Conclusion}

We considered the models of neutron stars in $f(R)=R+\alpha R^2$
gravity and its equivalent scalar-tensor theory in Einstein frame.
For neutron star matter the realistic equations were used. In
General Relativity for Tolman-Oppenheimer-Volkoff equations we
have Schwarzschild solution outside the star but for modified
gravity situation is more complex. One need to integrate equations
outside the star with zero energy density and pressure and check
consistency with asymptotic flatness on spatial infinity. As
result the gravitational sphere surrounding star appears with some
contribution to gravitational mass from viewpoint of distant
observer. For negative values of $\alpha$ the careful
consideration shows that gravitational mass grows infinitely with
distance. Although we have for scalar curvature required
asymptotic $R\rightarrow\infty$ at $r\rightarrow\infty$ but the
metric function $B(r)\rightarrow 1$ more slowly then $r^{-1}$ and
therefore function $m(r)$ grows with $r$. Another situation take
place for $\alpha>0$. The mass confined by stellar surface
decreases in comparison with General Relativity with increasing of
$\alpha$ (for given central density). But the net effect due to
gravitational sphere is increasing of gravitational mass. This
effect doesn't depend from equation of state for dense matter. For
equivalent scalar-tensor theory with massless dilaton scalar field
we have another interpretation. Stellar mass grows with $\alpha$
but this occurs due to increasing mass within stellar surface. The
contribution of dilaton sphere to gravitational mass is
negligible. On spatial infinity the solutions for mass profiles
coincide.

We considered the possible influence of deviation from general theory of
relativity on coalescence of neutron stars. We assumed that
effectively this influence can be regarded as modification of EoS
for dense matter and parameters of neutron stars consequently.
This assumption allows to conclude that in a case of scalar-tensor
theory with potential $V(\phi)\sim (1-e^{-2\phi/\sqrt{3}})^{2}$
threshold mass increases. In future one can hope that EoS for dense matter will be established with sufficient accuracy and therefore statistics of events like GW170817 can help us
discriminate between GTR and simple $R$-square gravity.


\begin{thebibliography}{99}

    \bibitem{Capozziello2002} S. Capozziello, Int. J. Mod. Phys. {\bf D 11}, 483 (2002).

    \bibitem{Capozziello2003} S. Capozziello, S. Carloni, A. Troisi, Recent Res. Dev. Astron. Astrophys. {\bf 1}, 625 (2003).

    \bibitem{Odintsov2003} S. Nojiri, S.D. Odintsov, Phys. Rev. {\bf D68}, 123512
    (2003).

    \bibitem{Carroll2004} S.M. Carroll, V. Duvvuri, M. Trodden and M.S. Turner, Phys. Rev. {\bf D 70}, 043528 (2004).

    \bibitem{Odintsov2011} S. Nojiri and S.D. Odintsov, Phys. Rept. {\bf 505}, 59 (2011).

    \bibitem{Capozziello2010} S. Capozziello and V. Faraoni, ``Beyond Einstein Gravity'', New York, Springer (2010).

    \bibitem{Capozziello2011} S. Capozziello and M. De Laurentis, Phys. Rept.{\bf  509}, 167 (2011).

    \bibitem{Cruz2012} A. de la Cruz-Dombriz and D. Saez-Gomez, Entropy {\bf 14}, 1717 (2012).

    \bibitem{Demianski2006} M. Demianski et al., Astron. Astrophys. {\bf 454}, 55 (2006).

    \bibitem{Perrotta2000} V. Perrotta, C. Baccagalupi, S. Matarrese, Phys. Rev. {\bf D 61}, 023507 (2000).

    \bibitem{Hwang2001} J.C. Hwang, H. Noh, Phys. Lett. {\bf B 506}, 13 (2001).

    \bibitem{Psaltis2008} D. Psaltis, Living Reviews in Relativity, {\bf 11}, 9 (2008).

    \bibitem{Demorest} P. B. Demorest, T. Pennucci, S. M. Ransom, M. S. E. Roberts,
and J.W. T. Hessels, Nature (London) 467, 1081 (2010).

\bibitem{Antoniadis} J. Antoniadis, P. C. C. Freire, N. Wex, T. M. Tauris, R. S.
Lynch et al., Science 340, 1233232 (2013).

\bibitem{Ozel} Ozel, F., Baym, G., Guver, T. 2010, Phys. Rev., D82,
101301.

\bibitem{Suleimanov} Suleimanov, V., Poutanen, J., Revnivtsev, M., Werner, K.
2011, Astrophys. J., 742, 122.

\bibitem{Lattimer} Lattimer, J. M., Lim, Y. 2013, Astrophys. J., 771,
51.

\bibitem{Guver} Guver, T., Ozel, F. 2013, Astrophys. J., 765,
L1.

\bibitem{Lattimer-14} Lattimer, J. M., Steiner, A. W. 2014, Astrophys. J., 784,
123.

\bibitem{Ozel-15} Ozel, F., Psaltis, D. 2015, Astrophys. J., 810,
135.

\bibitem{Ozel-16} Ozel, F., Freire, P. 2016, Annu. Rev. Astron. Astrophys., 54,
401.

\bibitem{Ozel-16-2} Ozel, F., Psaltis, D., Guver, T., et al. 2016, Astrophys. J.,
820, 28


\bibitem{Lattimer-04} Lattimer, J. M., Prakash, M. 2004, Science, 304, 536.

\bibitem{Lattimer-16} Lattimer, J. M., Prakash, M. 2016, Phys. Rept., 621,
127.

  \bibitem{Cooney2010} A. Cooney, S. De Deo and D. Psaltis, Phys. Rev. {\bf D 82}, 064033 (2010).

\bibitem{Arapoglu2011} S. Arapoglu, C. Deliduman, K.Y. Ek\c{s}i, JCAP {\bf 1107}, 020 (2011).

\bibitem{Alavirad2013} H. Alavirad, J.M. Weller, arXiv:1307.7977 [gr-qc].

\bibitem{Astashenok2013} A. Astashenok, S. Capozziello, S. Odintsov, JCAP \textbf{12}, 040 (2013).

\bibitem{Astashenok2014} A. Astashenok, S. Capoziello, S. Odintsov, Phys. Rev. D 89, 103509 (2014) arXiv:1401.4546 [gr-qc].

\bibitem{Cheoun2013} M.-K. Cheoun et al., arXiv:1304.1871 [astro-ph.HE].

\bibitem{Astashenok2015} A.V. Astashenok, S.Capozziello, S.D. Odintsov, Astrophys. Space Sci. 355, 333 (2015) arXiv: 1405.6663 [gr-qc].

\bibitem{Astashenok2015} A. Astashenok, S. Capozziello, S. Odintsov, Phys. Lett. B \textbf{742}, 160 (2015).

\bibitem{Kokkotas} K.V. Staykov, D.D. Doneva, S.S. Yazadjiev, K.D.
Kokkotas, JCAP 1406, 003 (2014).

\bibitem{Kokkotas-1} K.V. Staykov, D.D. Doneva, S.S. Yazadjiev, K.D.
Kokkotas, arXiv:1407.2180 [gr-qc].

\bibitem{APR} Akmal, A. and Pandharipande, V.~R. and Ravenhall, D.~G.,
Phys. Rev. C 58, 1804 (1998).

\bibitem{SLy} Chabanat, E. and Bonche, P. and Haensel, P. and Meyer,
J. and Schaeffer, R., Nucl. Phys. A635, 231 (1998).

\bibitem{SLy-4} Douchin, F. and Haensel, P., Astr. Ap. 380, 151 (2001).

\bibitem{GM} Glendenning, N.~K. and Moszkowski, S.~A., Phys. Rev. Lett.
67, 2414 (1991).

\bibitem{MYN} T. Miyatsu, S. Yamamuro, K. Nakazato, arXiv:1308.6121v1 [astro-ph.HE] (2013).

\bibitem{Resco2016} M. A. Rescoa, A. de la Cruz-Dombrizb, F. J. L. Estradaa, V. Z. Castrilloa, arXiv:1602.03880v2 [gr-qc].

\bibitem{LIGO} LIGO Scientific Collaboration, Virgo Collaboration
et al. ApJL 848, L12 (2017).

\bibitem{Abbott} B.P. Abbott et al. Astrophys. J., 848, L13 (2017)
arXiv:1710.05834[astro-ph.HE].

\bibitem{Abbott-2} B.P. Abbott et al., Phys. Rev. Lett., 161101, 119
(2017) arXiv:1710.05832[astro-ph.HE].

\bibitem{Abbott-3} B.P. Abbott et al., Astrophys. J., 848, L12
(2017) arXiv:1710.05833[astro-ph.HE].

\bibitem{Margalit} B. Margalit, B. Metzger, arXiv:1710.05938v2
[astro-ph.HE].

\bibitem{Bauswein} A. Bauswein, O. Just, H.-T. Janka, N.
Stergioulas, arXiv:1710.06843v2[astro-ph.HE].

\bibitem{Kasen} D. Kasen, B. Metzger, J. Barnes, E. Quataert,
E. Ramirez-Ruiz, Nature, 2017, arxiv:1710.05463[astro-ph.HE].

\bibitem{Bauswein-2} A. Bauswein, T.W. Baumgarte, H.-T. Janka, Phys. Rev. Lett. 111, 131101 (2013).


\bibitem{Cowperthwaite} Cowperthwaite, P. S., Berger, E., Villar, V. A., et al. 2017,
ApJL, 848, L17.

\bibitem{Chornock} Chornock, R., Berger, E., Kasen, D., et al. 2017, ApJL, 848,
L18.

\end{thebibliography}
\end{document}